# Efficient and scalable scheme for overcoming the pulse energy bottleneck of single-cycle laser sources

Lu Xu,[1,2,4] and Eiji J. Takahashi[1,3, *]

[1]Ultrafast Coherent Soft X-ray Photonics Research Team, RIKEN Center for Advanced Photonics, RIKEN, 2-1 Hirosawa, Wako 3510198, Japan
[2]Attosecond Science Research Team, RIKEN Center for Advanced Photonics, RIKEN, 2-1 Hirosawa, Wako 3510198, Japan
[3]Extreme Laser Science Laboratory, RIKEN Cluster for Pioneering Research, RIKEN, 2-1 Hirosawa, Wako 3510198, Japan
[4]e-mail: lu.xu@riken.jp
*Corresponding author: ejtak@riken.jp

**We propose a novel scheme called advanced dual-chirped optical parametric amplification (DC-OPA) that employs two kinds of nonlinear crystals ($BiB_3O_6$ and MgO-doped lithium niobate) to overcome the bottleneck of pulse energy scalability for single-cycle mid-infrared (MIR) laser pulses. In experiments, the advanced DC-OPA scheme achieved carrier-to-envelope phase-stable MIR laser pulses for a bandwidth of over one octave (1.4-3.1 μm) with an output pulse energy of 53 mJ. The pulse duration was compressed to 8.58 fs, which corresponds to 1.05 cycles with a central wavelength of 2.44 μm and peak power of 6 TW. To our knowledge, the obtained values for the pulse energy and peak power are the highest achieved for optical parametric amplification of single-cycle MIR laser pulses. Thanks to the energy scalability of the advanced DC-OPA scheme, it is potentially applicable to multi-TW sub-cycle laser pulses.**

## 1. Introduction

In the last decades, breakthroughs in the generation of high-energy and few-cycle infrared (IR) or mid-IR (MIR) lasers has improved understanding on the interaction of intense laser with matter, which has led to developments in a wide range of fields such as coherent X-ray high-order harmonic generation (HHG) [1], laser plasma accelerator [2], time-resolved imaging of molecular structures [3], sub-femtosecond-scale electron emission [4], and wave-packet dynamics in atoms/molecules [5]. Ultrafast laser pulses stable in the carrier-to-envelope phase (CEP) can be used to isolate the electron dynamics in the strong-field interactions, where the duration of the laser pulse is the key parameter that determines the characteristics of strong-field processes. Recently, high-energy single-cycle or even sub-cycle IR laser pulse have been drawing attention in the field of ultrashort optics. For example, research on HHG phenomena requires few-cycle laser pulses because the "cut-off" law deduced in the semi-classical three-step model [6] has shown that the cutoff photon energy of a high-order harmonic spectrum extends quadratically with the wavelength of the driving laser. Isolated attosecond pulse (IAP) emission, where the electron recombination in the HHG process is confined to a half-cycle, requires both a sufficiently short pulse duration consisting of a few-cycle electric field and CEP stability of the driving pulse. Currently, IAP emission driven by a few-cycle IR laser source has been widely demonstrated to have a limited energy output on the scale of picojoules, particularly in the water window region [7-12]. The most direct and effective way to overcome the severe influence of HHG photon flux scaling [13] and increase the output photon flux of IAP in the kiloelectronvolt region is to increase the pulse energy of the driving IR laser source. Thus, the development of a high-energy, single-cycle and CEP-stable laser source with a long wavelength is strongly desirable to significantly scale up photon energy, photon flux, and continuum bandwidth for IAP, not only in attosecond science but also in strong-field physics [2-5, 14, 15].

One approach to developing energetic few-cycle IR/MIR laser pulses is post-compression [16], in which high-average-power laser pulses from a conventional amplifier chain are compressed to few-cycle via self-phase modulation (SPM) in gases/solids [17-25]. Elu et al. [26] used a gas-filled antiresonant-guiding photonic crystal fiber to achieve the lowest cycle number of 1.35 cycles at a central wavelength of 3.25 μm but with a pulse energy of 60 μJ. Fan et al. [27] used a gas-filled hollow-core fiber to achieve the highest output pulse energy of 40 mJ, where the input pulse energy and pulse duration were 70 mJ and 220 fs, respectively. However, the output temporal duration was 25 fs at 1 μm, which is not sufficient for few-cycle. Currently, to obtain a few-cycle pulse output with high pulse energy, the cascaded post-compression scheme was demonstrated by Tsai et al. [28] with a pulse energy of 0.98 mJ and pulse duration of 3.1 fs (1.05 cycle at 885 nm), which results a perk power of 315 GW. Fibich and Gaeta [29] showed that the critical power of self-focusing, which is a limit to ensure that the beam does not collapse during SPM, gives a rough estimation of the power scaling with post-compression. For example, helium is commonly employed for high-energy post-compression with a gas pressure of ~ 1 atm. At a wavelength of 2 μm, the critical power would be approximately 2 TW, which indicates that post-compression is difficult to achieve a peak power beyond multiple terawatts with a few-cycle pulse duration in principle. In actual experiments, the beam quality is a key parameter for the stable propagation of laser pulses via SPM. Notably, the post-compression is not an amplification method for laser pulses. Therefore, it is not an ideal scheme to scale up the output energy of a single-cycle pulse.

Optical parametric amplification (OPA) is another approach that can generate femtosecond IR/MIR laser pulses with wide frequency tunability [30-35]. Brida et al. [30] employed a straightforward OPA scheme and reported a pulse duration of sub-two cycles centered at 1.6 μm with a pulse energy of 3 μJ. However, the damage threshold of nonlinear crystals limits the energy scalability of OPA at larger pulse energies. Various schemes have been proposed to circumvent the damage to the nonlinear crystals due to a high-energy pump laser, such as optical parametric chirp pulse amplification (OPCPA) [36-40], frequency domain optical parametric amplification [41] and dual-chirped optical parametric amplification (DC-OPA) [42-47]. However, even though DC-OPA has been used to extend the pulse energy of IR/MIR laser pulses to tens or even hundreds of millijoules [47], the shortest output pulse duration is still limited to sub-two cycles because of the gain bandwidth limit of a single nonlinear crystal. Parallel waveform synthesis with OPA, which coherently combines CEP-stable pulses emerging from different OPA chains, has been considered as an efficient technique to realizing tailored optical waveforms with sub-cycle pulse duration [48-51]. However, the pulse energy scalability is limited by the complexity of the laser system, and the highest reported output pulse energy is less than a millijoule [51]. Currently, no methods have been demonstrated that can directly amplify the pulse energy of single-cycle or even sub-cycle laser pulses to multiple millijoules.

In this paper, we first propose a novel amplification method called advanced DC-OPA scheme to overcome the bottleneck of pulse energy scalability in a single-cycle IR/MIR laser system. The advanced DC-OPA scheme utilizes more than one nonlinear crystal and optimizes the phase-matching (PM) of each crystal and the chirp matching between the pump and seed laser pulses to support amplification with a bandwidth of over one octave. Based on the 10 Hz joule-class Ti:sapphire pump laser and the advanced DC-OPA scheme, where the $BiB_3O_6$ (BiBO) and MgO-doped lithium niobate (MgO:LiNbO$_3$) nonlinear crystals are combined in each stage of parametric amplifiers, an over one-octave bandwidth MIR pulses are amplified with the pulse energy of 53 mJ centered at 2.44 μm. After the pulse compression using a sapphire bulk, a temporal pulse duration is down to 8.58 fs, which corresponds to 1.05 cycle at 2.44 μm. Besides that, the measured shot-to-shot stability in single-shot CEP value is 228 mrad rms, and the evaluated $M^2$ values of focused beams in the horizontal and vertical directions are 1.24 and 1.29, respectively. Finally, we discuss the possibility to demonstrate the next-generation DC-OPA, which is a high-energy, sub-cycle, CEP-stable MIR laser source as the future prospect of the advanced DC-OPA scheme.

## 2. Method

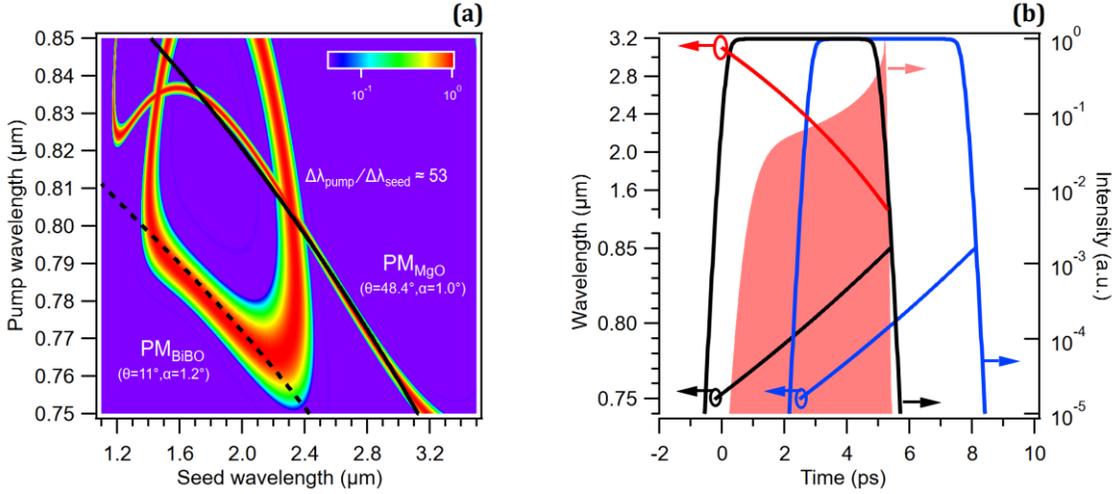

Fig. 1. Calculated PM efficiency as a function of the pump and seed laser wavelengths and the temporal overlap between the pump and seed laser pulses. (a) PM efficiencies of a type-I BiBO crystal and type-I MgO:LiNbO$_3$ crystals, where the black dashed and solid lines (slopes $\Delta\lambda_{pump}/\Delta\lambda_{seed}$ of 0.053 by linear fitting) represent the wavelengths of the chirped seed laser pulses and corresponding pump laser pulses for the BiBO and MgO:LiNbO$_3$ crystals, respectively. (b) Temporal overlap between the pump and seed laser pulses corresponding to the chirp matching in (a). The left and right axes represent the wavelengths and normalized peak powers, respectively, corresponding to each moment in the time domain (bottom axis). The red fill represents the intensity profile (right axis) of the chirped seed laser pulse in the time domain, and the red curve shows the change in wavelength (left axis) of the chirped seed laser pulse in time domain. The black curves show the intensity profile (right axis) and change in wavelength (left axis) of the chirped pump laser pulse in the time domain to realize the chirp matching condition in the MgO:LiNbO$_3$ crystal shown in Fig. 1(a). The blue curves show the intensity profile (right axis) and change in wavelength (left axis) of the chirped pump laser in the time domain to realize the chirp matching condition in the BiBO crystal shown in Fig. 1(a).

We previously demonstrated a DC-OPA scheme using a BiBO crystal with a bandwidth of 1.2–2.2 μm that achieved a pulse duration of sub-two cycles and peak power of multiple terawatts [47]. In this study, we extended the DC-OPA scheme to realize an amplified bandwidth of over one-octave. However, we first need to discuss the amplified bandwidth of the DC-OPA with a single nonlinear crystal. Fig. 1(a) plots the PM efficiency (i.e., sinc$^2(\Delta kL/2)$) as a function of the pump and seed laser wavelengths, where $\Delta k$ is the phase mismatch, L is the nonlinear crystal length (set to 5 mm here), and the pump laser pulses have a fixed spectral range of 0.75-0.85 μm corresponding to a typical chirp pulse amplification Ti:sapphire laser. The color contours represent the PM efficiency of a non-collinear geometric configuration employing both type-I BiBO and type-I MgO:LiNbO$_3$ crystals with PM angles θ of 11.0° and 48.4°, respectively, and non-collinear angles α of 1.2° and 1.0°, respectively. In the DC-OPA scheme, the amplified bandwidth is determined by the characteristics of the nonlinear crystals. Fig. 1(a) indicates that MgO:LiNbO$_3$ may be able to realize an amplified bandwidth of one octave (1.6-3.2 μm) under the linear chirped configuration. However, this ideal case is not easy to realize because the chirp configuration of the seed laser pulse in the experimental setup cannot be as linear as that of the pump laser pulse. The advanced DC-OPA scheme smoothly connects the PM regions of the two kinds of nonlinear crystals to achieve an amplified bandwidth of over one octave. We Considered the transmittance and gain bandwidth of commonly used nonlinear crystals and finally selected BiBO and MgO:LiNbO$_3$ crystals for our experimental demonstration of the amplification of an MIR laser source by the advanced DC-OPA scheme.

For the advanced DC-OPA scheme to achieve amplification with a broad bandwidth, the most critical aspect is chirp management of pump and seed laser pulses. Specifically, pump laser pulses, which are stretched using grating pairs, have an almost linear chirp. Meanwhile, seed laser pulses, which are stretched in the acousto-optic programmable dispersive filter (AOPDF) and then compressed in the bulk material, have a nonlinear chirp. Under the PM condition, Fig. 1(a) shows the one-to-one correspondence between the seed and pump laser wavelengths in the form of curves (i.e., solid and dashed black lines). These curves are linearly fitted to a slope ($\Delta\lambda_{pump}/\Delta\lambda_{seed}$) of 0.053 to intuitively show the chirp matching relationship between the pump and seed laser pulses. To fully utilize the pump energy and prevent damage to the nonlinear crystal, the laser pulses in the DC-OPA amplifiers need to be stretched by several picoseconds in advance. Therefore, we employed sapphire bulk with a length of 40 mm as a compressor to achieve a compression of several picoseconds. As shown in Fig.1 (a), the dashed black line indicates that the chirp matching between the seed laser pulse (bottom axis) and corresponding pump laser pulse (left axis) is well within the PM region of the BiBO crystal, which indicates broadband parametric amplification with a wavelength of 1.4-2.4 μm. The solid black line shifts the same chirp matching relationship to the PM region of the MgO:LiNbO$_3$ crystal and shows that the overlapping part can support parametric amplification with the wavelength of 2.2–3.1 μm. For chirped pulses, chirp matching is reflected not only by corresponding wavelengths in the frequency domain but also by the overlapping positions of the pump laser pulse and seed laser pulse in the time domain. While Fig. 1(a) shows the chirp matching in the frequency domain, Fig. 1(b) shows the same chirp matching relationship in the time domain. In summary, the advanced DC-OPA scheme combines the PM regions of type-I BiBO and type-I MgO:LiNbO$_3$ nonlinear crystals and optimizes the temporal overlap of the pump and seed laser pulses to realize an amplified bandwidth for MIR laser pulses of over one octave. The proposed scheme establishes a solid foundation for subsequent experimental research.

## 3. Results

Fig. 2 illustrates the experimental setup of a high-energy, single-cycle, and CEP-stable MIR laser system based on the proposed advanced DC-OPA scheme. Compared with the previously reported system [47], we made several major upgrades to obtain high-energy single-cycle MIR laser pulses.

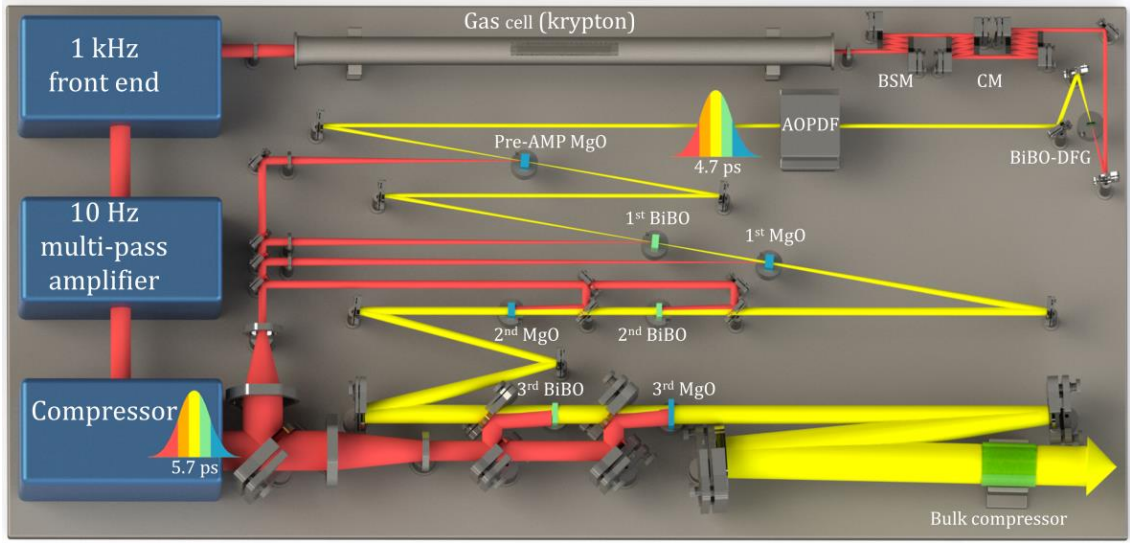

Fig. 2. High-energy, single-cycle MIR laser source based on the advanced DC-OPA scheme. BSM, band stop mirror; CM, chirped mirror; AOPDF, acousto-optic programmable dispersive filter; DFG, difference frequency generation.

Specifically, (i) the Ti:sapphire pulses with a total pulse energy of 750 mJ and pulse duration of ~5.7 ps (full width at $10^{-3}$, nearly linear negative chirp) were applied as the pump for the advanced DC-OPA. (ii) To further broaden the bandwidth of the seed laser, a 1-mm-thick BiBO crystal with a cutting angle of 60° for type-II PM in the x-z plane was employed in the difference frequency generation (DFG) process for CEP-stable laser pulse generation. The spectrum of the pump laser (0.8 μm) was broadened for the DFG process via optical filamentation in a 1.6-bar krypton gas cell [52], which was followed by a band-stop mirror for spectral reshaping [53] and a chirped mirror for dispersion compensation.

Finally, the passively CEP-stable, over one-octave MIR (1.4–3.1 μm) seed laser pulses were generated via DFG in Type-II BiBO crystal, whose spectrum was shown in Fig.3 (black filled profile). After stretched in the AOPDF (DAZZLER/HR45-1450-3000), the MIR seed laser pulses, with a dispersion predominantly given as the inverse of the sapphire bulk compressor and pulse duration of ~4.7 ps (full width at $10^{-3}$), were produced as the seed for the advanced DC-OPA. (iii) There were one pre-amplification stage using MgO:LiNbO$_3$ crystal and three post-amplification stages combining two kinds of nonlinear crystals (i.e., BiBO and MgO:LiNbO$_3$) in the advanced DC-OPA. We employed the pre-amplification stage to compensate for the absorption by the BiBO crystals in the subsequent amplifiers of wavelengths longer than 2.4 μm and the difference in quantum efficiency of different wavelengths during the parametric amplification process. In contrast to the pre-amplifier, the three subsequent post-amplifiers each contained both a BiBO crystal and MgO:LiNbO$_3$ crystal, and the BiBO crystal was in front of the MgO:LiNbO$_3$ crystal to reduce the effect of the absorption by BiBO on the final output pulse energy. We also installed a nitrogen-filled sealed box to cover the advanced DC-OPA module to avoid adsorption by water in the air, which causes significant spectral phase distortion in the compression of high-energy IR/MIR laser pulses [47].

**Table 1. Parameters in the single cycle laser system** [a]

| Stages \ Parameters | Pre-amp | First stage | Second stage | Third stage_1 | Third stage_2 |
|---|---|---|---|---|---|
| Pump energy (mJ) | 0.4$_{MgO}$ | 0.8$_{BiBO}$ 0.7$_{MgO}$ | 13$_{BiBO}$ 17$_{MgO}$ | 400$_{BiBO}$ 320$_{MgO}$ | 320$_{BiBO}$ 400$_{MgO}$ |
| Pump Int. (GW/cm$^2$) | 25$_{MgO}$ | 45$_{BiBO}$ 40$_{MgO}$ | 40$_{BiBO}$ 50$_{MgO}$ | 50$_{BiBO}$ 40$_{MgO}$ | 40$_{BiBO}$ 50$_{MgO}$ |
| Gain | ~100 | 2000 | 1600 | 77 | 66 |
| Output energy | ~0.25 nJ | 0.5 μJ | 0.8 mJ | 61 mJ | 53 mJ |

[a] The subscripts BiBO and MgO indicate parameters acting on the BiBO and MgO:LiNbO$_3$ crystals, respectively.

In the advanced DC-OPA module, the pre-amplifier and subsequent three post-amplifiers employed type-I MgO:LiNbO$_3$ crystals with thicknesses of 6, 6, 5, and 4 mm, respectively. The three post-amplifiers also employed type-I BiBO crystals with thicknesses of 5, 4, and 4 mm, respectively. Both the PM and non-collinear angles of the MgO:LiNbO$_3$ and BiBO crystals were consistent with Fig. 1(a). As shown in Fig.2, we employed focusing geometry for the interacting laser pulses in both the pre-amplifier and first post-amplifier. The nonlinear crystals were placed in the overlap of the Rayleigh lengths for the pump and seed beams to maintain good PM condition between the pump and seed lasers. In the second and third post-amplifiers, both the pump and seed laser pulses were collimated. Table 1 lists the parameters used in each amplification stage and the output performance of each amplifier.

We included some supporting functions in the amplification chain based on the advanced DC-OPA scheme to ensure the output quality of the MIR laser system. (i) As shown in Fig. 1(b), the time delays of the pump and seed laser pulses were inconsistent for BiBO and MgO:LiNbO$_3$ crystals, so the time delay corresponding to each crystal needed to be adjusted independently. (ii) To reduce the influence of the amplified parametric fluorescence, which particularly emerges under a high pump laser intensity, the pump laser pulses were separated into seven nonlinear crystals in total. (iii) Because the beam profile of the amplified beam is significantly dependent on that of the pump beam, the image of the pump beam was relayed to the surface of each nonlinear crystal to maintain a flattop-like profile for the pump beam. (iv) Finally, we added intensity holes around the wavelengths of 1.8 and 2.5 µm to the seed laser spectrum via AOPDF to avoid saturation around these wavelength regions. This greatly reduced the influence of spectral phase distortion caused by reversed energy flow in the parametric process on the following compression work [46].

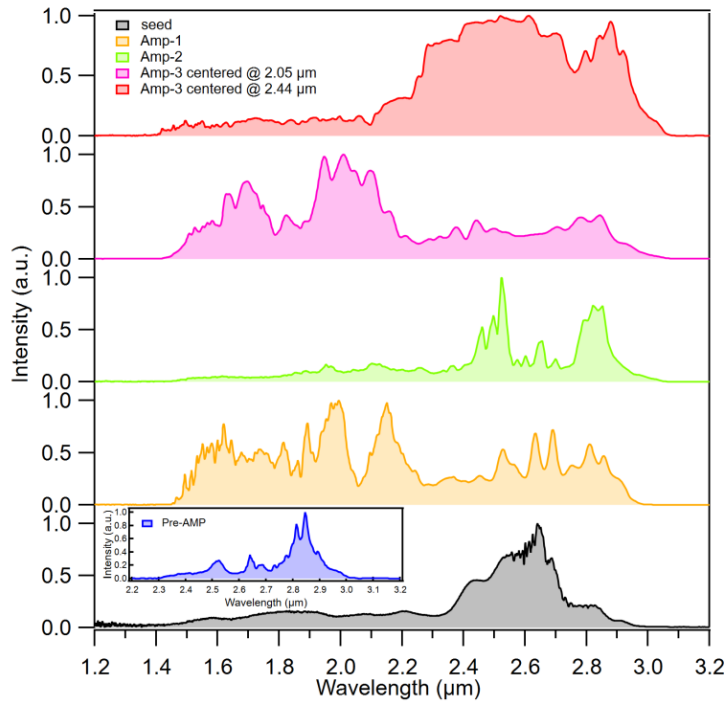

Fig. 3. Evolution of the seed laser spectrum: input seed spectrum for advanced DC-OPA (black solid line with filling), output of the pre-amplifier based on MgO:LiNbO$_3$ (blue solid line with filling), output of the first amplifier based on BiBO and MgO:LiNbO$_3$ (orange solid line with filling), the output of second amplifier based on BiBO and MgO:LiNbO$_3$ (green solid line with filling) and the output of third

amplifier based on BiBO and MgO:LiNbO$_3$ (pink solid line with filling and red solid line with filling for the central wavelength of 2.05 μm and 2.44 μm, respectively).

By adjusting the non-collinear angle and time delay between the pump and seed introduced in each nonlinear crystal, the combination of the pre-amplifier (MgO:LiNbO$_3$) and first post-amplifier (BiBO and MgO:LiNbO$_3$) realized a the relatively balanced amplification of the full spectrum at 1.4-3.1 μm (orange filled profile in Fig.3). In the second amplifier stage, the MgO:LiNbO$_3$ crystal was thicker than the BiBO crystal and was subjected to a higher pump power intensity, so the central wavelength of the amplified spectrum after the second stage was shifted toward the longer wavelength band (green filled profile in Fig.3). It is worth noting that, a higher pump power intensity was applied to the MgO:LiNbO$_3$ crystal than that to the BiBO crystal in the last amplifier (third stage_2 in Table 1), the MIR laser system obtained a final output pulse energy of 53 mJ with a total conversion efficiency from the pump laser to seed laser of 7%. Based on these optimizations, an over one-octave (1.4–3.1 μm) output bandwidth with the pulse energy of >50 mJ is attained based on the advanced DC-OPA scheme. To consider the number of cycles in a pulse envelope, we applied the center-of-gravity (COG) concept, which has been recognized to determine the carrier wavelength (i.e., central wavelength) of laser pulses [50], to define the central wavelength of the spectrum. According to the COG, the central wavelength of the output spectrum (red filled profile in Fig. 3) after the final amplifier stage was 2.44 μm. Owing to these optimizations, the advanced DC-OPA scheme output a pulse energy of >50 mJ for a bandwidth of over one octave (1.4–3.1 μm).

When the pump laser energies allocated to the two crystals in the last amplifier were exchanged (i.e., a higher pump laser energy was applied to the BiBO crystal (third stage_1 in Table 1)), the amplification energy of the final stage was increased to 61 mJ, but the central wavelength shifted to 2.05 μm (pink filled profile in Fig. 3). This is because a higher gain was achieved in the short-wavelength region than in the long-wavelength region with this arrangement. This demonstrates another advantage of the advanced DC-OPA scheme: the tunability of the central wavelength.

After the amplification, a 40-mm-long sapphire bulk compressor was employed to temporally compress the MIR laser pulses with the diameter of the beam expanded to 60 mm (the calculated B-integral was 0.11). To characterize the temporal duration of the compressed pulses with an over one-octave bandwidth, third-order harmonic generation frequency-resolved optical gating (THG-FROG) was adopted for the MIR laser pulses. To fully compress the MIR laser pulses, the dispersion loaded in the AOPDF should be strictly optimized. Thus, the spectral phases of the measured laser pulses, that were retrieved from THG-FROG traces, were fed back to the AOPDF for the modification of dispersion compensation. After several iterations to correct the preset spectral phase in the AOPDF, fully compressed MIR laser pulses with a central wavelength of 2.44 μm were obtained with a nearly flat spectral phase, which was demonstrated in Fig. 4. The spectral intensity and spectral phase, which were retrieved from the measured THG-FROG traces expressed in Fig. 4(a), were demonstrated by the red solid line and black dashed line, respectively, in Fig. 4(d). The coincidence of the reconstructed spectrum (red solid line) and measured spectrum (black solid line) in terms of the range and trend of the wavelengths in Fig. 4(d) further proved that complete compression with a full bandwidth was achieved. The corresponding temporal pulse profiles of the retrieved pulses with a pulse duration of 8.58 fs (FWHM) and a transform-limited (TL) pulse duration of 8.22 fs were illustrated in Fig. 4(c), which corresponded to a 1.05-cycle duration of the electric field at the central wavelength of 2.44 μm and resulted in peak power of 6 TW. In addition, the output MIR laser pulses with a pulse energy of 61 mJ and the central wavelength of 2.05 μm were also measured via THG-FROG, in which the temporal pulse profiles with a pulse duration of 8.75 fs and a TL pulse duration of 8.18 fs were achieved. it indicated that the MIR laser pulses had a pulse duration of 1.28 cycles and peak power of 7 TW.

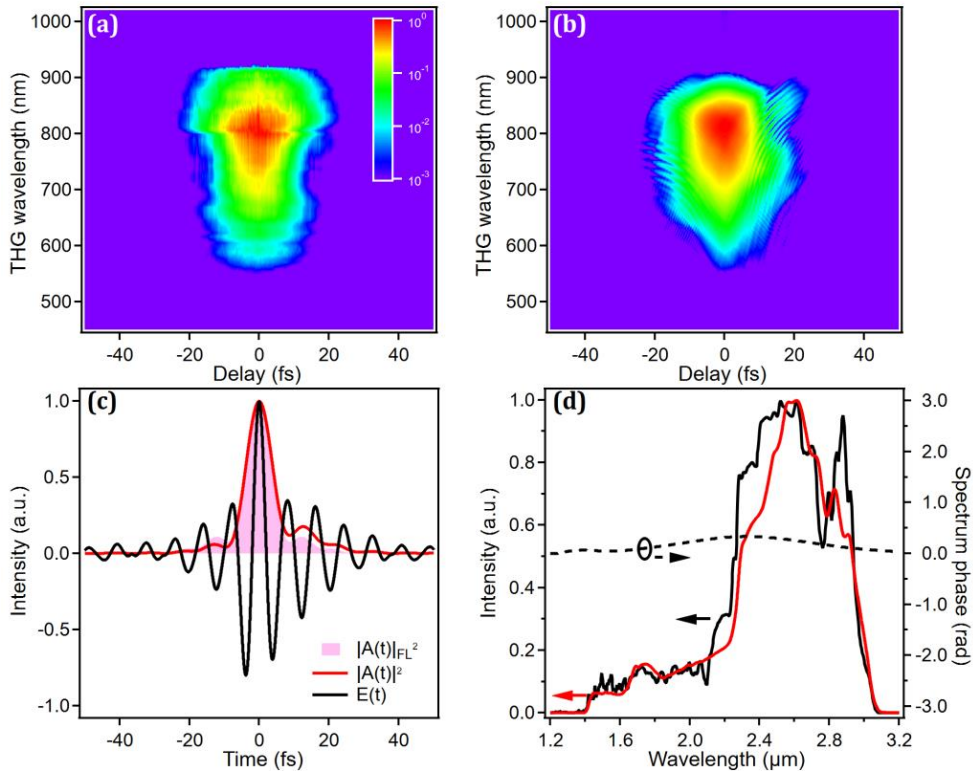

Fig. 4. THG-FROG results of the MIR laser pulses based on the advanced DC-OPA scheme. The measured THG-FROG traces with central wavelength of 2.44 µm is shown in (a). (b) The retrieved THG-FROG traces corresponding to (a) with an error of 0.78%. (c) Reconstructed temporal profile with a pulse duration of 8.58 fs (FWHM), where the electric field profile (black solid line) under the cos-like waveform (carrier wavelength, 2.44 µm) and its envelope squared (red solid line, transform-limited in pink filling) are shown respectively. (d) Reconstructed spectrum (red solid line) and spectral phase (black dash line), and measured spectrum of the final output centered at 2.44 µm (black solid line).

Subsequently, a home-built single-short f-to-2f interferometer was implemented after the sapphire bulk compressor to characterize the CEP stability of the MIR laser pulse. Because the spectral bandwidth of the compressed MIR laser pulses was over one octave, the longer wavelength components of the compressed pulses were frequency doubled by a $LiIO_3$ crystal and interfered with the shorter wavelength components through a polarizer to generate the interference fringe. The time evolution of the f–2f interference fringes, which were in the wavelength region of 1.45-1.48 µm, were shown in Fig. 5(a). The CEP values extracted from the spectrogram in Fig. 5(a), which included 1800 consecutive records with single-short, were demonstrated in Fig. 5(b) with an RMS of 228 mrad for 3 minutes.

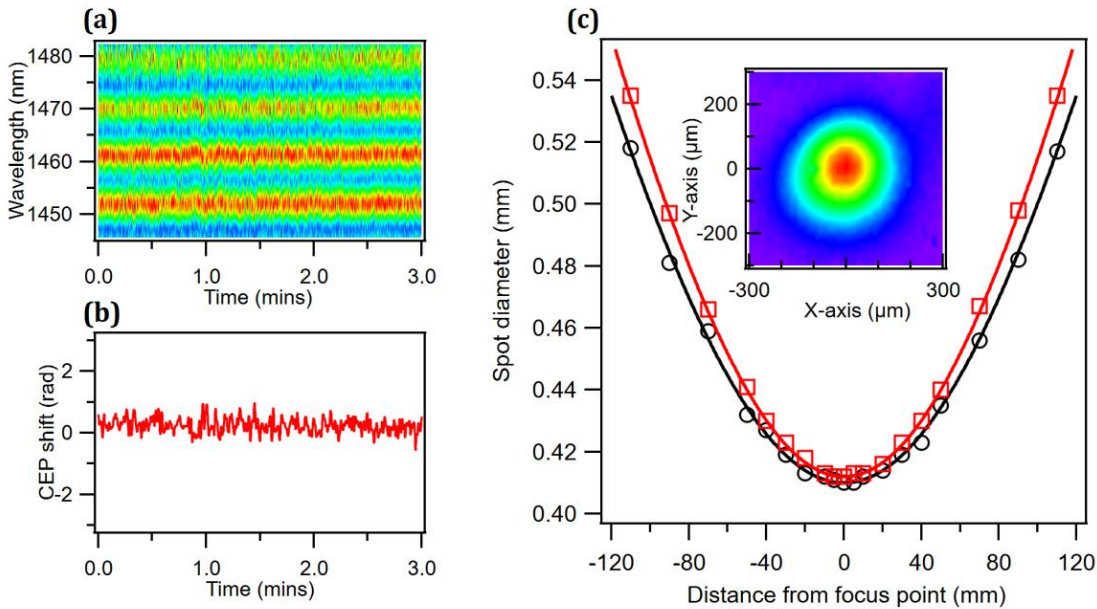

Fig. 5. Measurements of the single-short CEP in the f-to-2f interferometer and the beam quality. (a) Temporal evolution of the f–to-2f spectral interference fringes, where a single-shot of the spectrum was obtained for each record. (b) CEP shift extracted from the spectrogram depicted in (a). The rms error of the CEP shift was evaluated to be 228 mrad. (c) Evolution of the beam diameters (half width at $e^{-2}$ maximum) in the x-axis (horizontal direction, black circles) and Y-axis (vertical direction, red squares) with the beam propagation across the focal point. The evaluated $M^2$ values for X-axis and Y-axis by fitting each measured diameter were 1.24 (black solid line) and 1.29 (red solid line), respectively. The corresponding beam profile at the focal point is shown in the inset of (c).

Meanwhile, the quality of the focus beam profile is one of the great concerns in the strong-field experiments. A 0.5-m focusing geometry was used to verify the focusability of the output MIR laser beam. The measured beam profile with a diameter of ~410 μm (full width at $1/e^2$) on the focus plane was shown in the inset of Fig. 5(c). Concurrently, the diameter evolutions of the focusing beam profile up to 2 times of Rayleigh length across the focal point were recorded in the horizontal (black circles in Fig. 5c) and vertical directions (red squares in Fig. 5c), respectively. In each measurement, the diameters for the beam were calculated by fitting the measured profiles to a Gaussian function. Finally, the obtained $M^2$ (defined at the central wavelength of 2.44 μm) values of the focus beam profile were 1.24 and 1.29 in the horizontal and vertical directions, respectively.

## 4. Prospects and conclusion

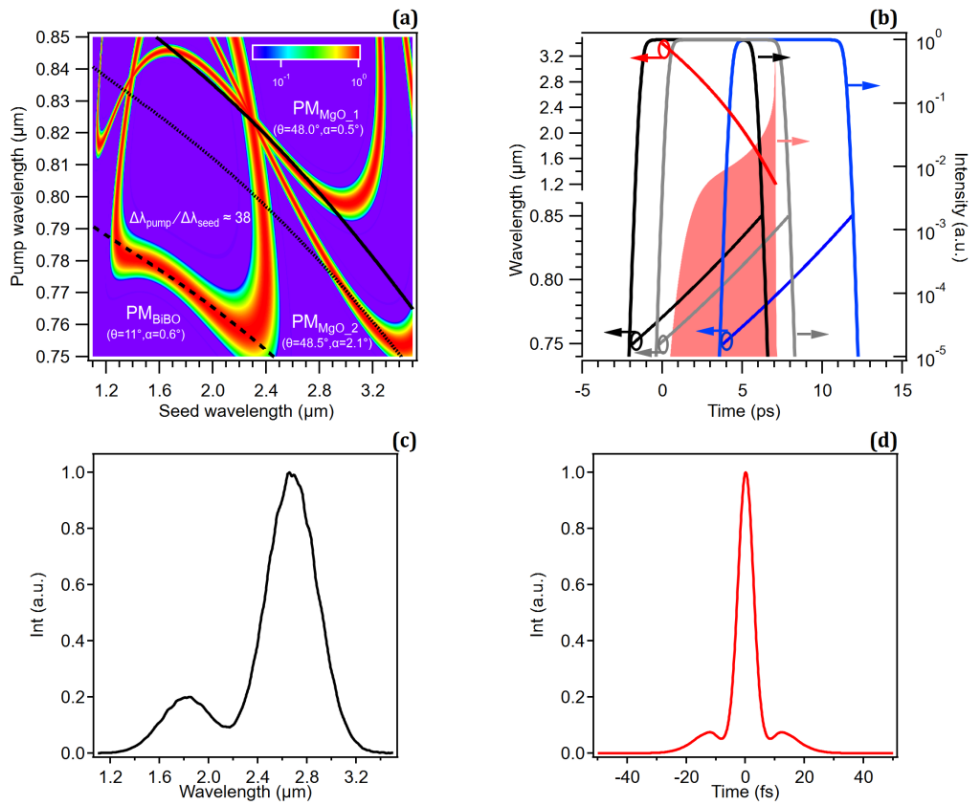

Fig. 6. Calculated PM efficiency as a function of the pump and seed laser wavelengths and the overlap between the pump and seed laser pulses in the time domain. (a) PM efficiency combining the type-I BiBO and type-I MgO:LiNbO$_3$ crystals. The black dashed, solid and dotted lines with slopes ($\Delta\lambda_{pump}/\Delta\lambda_{seed}$) of 0.038 by linear fitting represent the wavelengths of the chirped seed laser pulse and corresponding pump laser pulse in the BiBO and MgO:LiNbO$_3$ crystals, respectively. (b) Temporal overlap between the pump and seed laser pulses corresponding to the chirp matching in (a). The left and right axes represent the wavelengths and normalized peak powers, respectively, corresponding to each moment in the time domain (bottom axis). The red fill shows the intensity profile (right

axis) of the chirped seed laser pulse in the time domain. The red curve shows the change in wavelength (left axis) of the chirped seed laser pulse in the time domain. The black curves show the intensity profile (right axis) and change in wavelength (left axis) of the chirped pump laser pulse in the time domain to realize the chirp matching condition for the first MgO:LiNbO$_3$ crystal shown in Fig. 6(a). The gray curves show the intensity profile (right axis) and change in wavelength (left axis) of the chirped pulse laser pulse in the time domain to realize the chirp matching condition for the second MgO:LiNbO$_3$ crystal shown in Fig. 6(a). The blue curves show the intensity profile (right axis) and change in wavelength (left axis) of the chirped pump laser pulse in the time domain to realize the chirp matching condition for the BiBO crystal shown in Fig. 6(a). (c) Simulated final output spectrum of the designed sub-cycle MIR laser system. (d) Corresponding transform-limited pulse duration of the spectrum in (c).

Thanks to the various types and numbers of nonlinear crystals combined in the advanced DC-OPA scheme, one of the obvious advantages is its excellent energy scalability. Specifically, (i) In this paper, since the Ti:Sapphire laser was used as the pump of the advanced DC-OPA scheme, the optimal combination scheme of BiBO and MgO:LiNbO$_3$ crystals was finally adopted. For pump lasers with other wavelength regions, similar crystal combination methods (crystal types are not limited to BiBO and MgO:LiNbO$_3$) can be used to achieve ultra-broadband parametric amplification. (ii) In this study, we used a two-crystal combination mode, which was mainly limited by the bandwidth of the AOPDF and total pump laser energy in the experiment. Notably, based on a wider-bandwidth dispersion control device and higher total pump energy, a combination of multiple crystals can be used to achieve a broader-bandwidth parametric amplification and obtain a sub-cycle pulse duration. As the future prospects, we discuss the advanced DC-OPA scheme to amplify the sub-cycle MIR laser pulses based on the experimental results of a single-cycle, where the longest wavelength of the amplified MIR laser has reached the limitation of the transmission in BiBO as 3.4 μm.

As the PM efficiency in Fig. 6(a), where the non-collinear geometric configure is adopted in the combination of one BiBO crystal and two MgO:LiNbO$_3$ crystals. The PM angles θ are 11.0°, 48.0°, and 48.5°, which corresponds to non-collinear angles α of 0.6°, 0.5°, and 2.1°, respectively. The seed laser pulses with wavelength range of 1.2-3.4 μm can be successfully amplified based on the advanced DC-OPA scheme. To maintain the broadband amplification over nearly two octaves, the dispersion management of the pump and seed laser pulses is extremely critical. The chirp matching between the pump and seed laser pulses (black dashed line in Fig. 6(a)) well overlaps with the PM region of the BiBO crystal in the wavelength range of 1.2-2.4 μm. When the same chirp matching relationship ($\Delta\lambda_{pump}/\Delta\lambda_{seed} \approx 0.038$) is applied to the PM regions of the two MgO:LiNbO$_3$ crystals, the overlapping areas can effectively support parametric amplification in the wavelength bands of 2.2-2.9 μm and 2.8-3.4 μm, respectively. Correspondingly, the temporal overlapping between the seed laser pulse and three pump laser pulses is shown in Fig. 6(b). It indicates that the MIR laser pulses with a bandwidth of 1.8 octaves are fully amplified by the advanced DC-OPA scheme. Compared with the single-cycle experiments, the designed amplifier has the following differences. (i) two MgO:LiNbO$_3$ crystals are used in the pre-amplification to compensate for the difference in subsequent absorption loss and quantum efficiency; (ii) There are three crystals (BiBO + MgO:LiNbO$_3$ +MgO:LiNbO$_3$) in each post-amplification stage to ensure effective amplification of the full bandwidth. Based on the numerical simulation model [46] and Ti:sapphire pump laser pulses with a total pulse energy of over 1 J, the calculated final output spectrum and corresponding TL pulse duration were shown in Fig.6(c) and Fig.6 (d), respectively. The simulated output pulse energy reached 65 mJ with a total conversion efficiency of approximately 7 %. It indicated that the MIR laser pulses with a bandwidth of 1.8 octaves were fully amplified by the advanced DC-OPA scheme, which supported the TL pulse duration (6.2 fs) of 0.75 cycles at 2.47 μm.

To our knowledge, it is the first time to demonstrate an amplification method for single-cycle laser pulses. Based on the optimized experimental setup of the advanced DC-OPA scheme, where the combination of type-I BiBO and type-I MgO:LiNbO$_3$ nonlinear crystals was employed in the parametric amplification, an over 50 mJ, single-cycle (8.58 fs at 2.44 μm), 10 Hz, CEP-stable (228 mrad rms) MIR laser source was demonstrated, resulting in peak power of >6 TW. Furthermore, the prospect of a high-energy MIR laser source toward sub-cycle pulse duration was discussed by extending the combination of nonlinear crystals in the advanced DC-OPA scheme. In strong-field physics dealing with the interaction of laser with matter, the "cycle number of the electric field" within a pulse envelope plays an extremely important role. Especially, the cycle number significantly affects the continuous range of the HHG spectrum. When the cycle number of the driving laser pulses is 2, 1, and 0.7 within a Gaussian envelope, the corresponding percentages of the continuum region in the HHG spectrum are approximately 15%, 50%, and 73%, respectively – i.e., the continuum bandwidth reaches 500 eV at the 1-keV cutoff photon energy pumped by a single-cycle laser pulse. The less-cycle MIR laser source based on the advanced DC-OPA scheme will present the potential for the generation of x-ray with a duration of several attoseconds (even zeptosecond), which would offer breakthrough progress in the electron dynamics and correlations. Moreover, thanks to the excellent energy scalability of this advanced DC-OPA scheme, it is reasonable to obtain laser pulses with higher pulse energy and less cycle number of pulse duration based on different crystal combinations and a higher pump energy. The expansion of pulse energy can facilitate high flux detection conditions for research in strong field physics. Meanwhile, sub-cycle (even half-cycle) laser pulses can effectively avoid the superposition of phenomena caused by periodic oscillations of the laser field in the laser-matter interaction under strong field conditions, which can make the research results more intuitive and accurate.

**Funding.** Ministry of Education, Culture, Sports, Science and Technology of Japan (MEXT) through Grants-in-Aid under Grant Nos. 21H01850, in part, by the MEXT Quantum Leap Flagship Program (Q-LEAP) (Grant No. JP-MXS0118068681).

**Acknowledgments.** We thank Dr. Y. Nabekawa for the useful comments on measurement of the MIR pulse duration, Dr. B. Xue for support with THG-FROG, and Dr. Y.-C Lin for the discussion on COG. L. X. acknowledges support from MEXT Q-LEAP.